# An ultra compact on-beam $^3$He polarizer for separation of incoherent background in high Q-range SANS

E. Babcock[1], A. Ioffe[1], Z. Sahli[1]

[1] *Juelich Centre for Neutron Science at FRM 2, Forschungszentrum Juelich GmbH, 85747 Garching, Germany.*

Incoherent background can create an intrinsic problem for standard small angle neutron scattering measurements. Biological samples contain hydrogen which is a strong incoherent scatterer thus creating an intrinsic source of background that makes determination of the coherent scattering parameters difficult in special situations. This can especially be true for the Q-range from around 0.1-0.5 Å$^{-1}$ where improper knowledge of the background level can lead to ambiguity in determination of the samples structure parameters. Polarization analysis is a way of removing this ambiguity by allowing one to distinguish the coherent from incoherent scattering, even when the coherent scattering is only a small fraction of the total scattered intensity. $^3$He spin filters are ideal for accomplishing this task because they permit the analysis of large area and large divergence scattered neutron beams without adding to detector background or changing the prorogation of the scatter neutron beam. This rapid note describes a design for a new ultra compact SEOP based $^3$He polarized that can be used for this application. This is a work in progress, the magnetic system has been completed and tested. We are in the process of constructing the mechanical components for online polarization, such as a compact laser system and the oven described in this intermediate document.

Our system is based on a 17 cm long mu-metal solenoid. Clearly capped mu-metal solenoids can be used to create very uniform magnetic fields. However for neutron applications there must be holes in the caps to allow access for the scattered neutron beam. The holes undoubtedly create field gradients in the cavity. We have used an approach that combines partial shielding with mu-metal fingers, and compensation of the holes in the end caps with additional coils. Diagrams of the system are shown in the figures below.

A 1x1 cm$^2$ beam is assumed to be incident on the sample which would be a typical soft matter sample located in the brown holder directly before the right end cap. The cell is placed within 2 cm of the 1.5x1.5 cm$^2$ hole in this endcap. After passing through the continuously polarized $^3$He cell of D=6cm and L=5cm a square cross section beam can pass the partially shielded opening in the end cap on the left of the drawing.

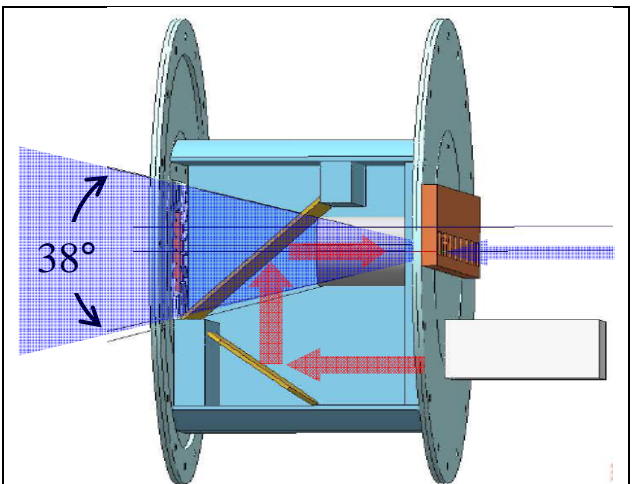

Fig. 1. Cut-away drawing of the magnetic cavity showing the optical access (red arrows) and the neutron access, violet.

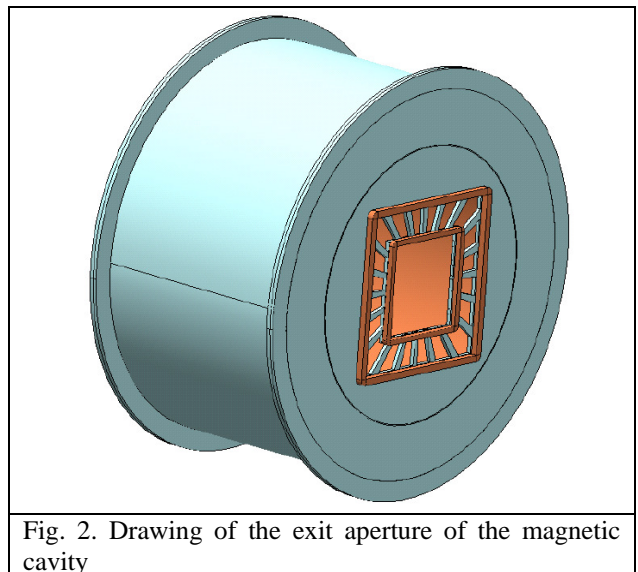

Fig. 2. Drawing of the exit aperture of the magnetic cavity

The laser head is placed outside of the magnetic cavity and the laser beam enters through a small hole in the right endcap.

Expansion of the laser beam to D=6 cm will be done with optic inside the magnetic cavity and the expanded beam will be reflected longitudinally along the neutron bema via a gold coated Si-wafer as shown in Fig 1.

As can be seen in Fig. 2 the exit aperture of the magnetic cavity is partially shielded by mu-metal fingers, which while reducing the effective open area of the hole they do not reduce the maximum angular coverage, and since for this application the scattering is isotropic the fingers will only cause a manageable loss in count rate from the blocked area.

Without these fingers the FEM calculations of the magnetic field homogeneity did not provide what we considered to be acceptable magnetic field gradient performance. It can also be seen in Fig. 2 that the inner and outer portions of the square hole in the exit endcap are compensated for the loss in magnetic flux from the holes with two square coils. The inner square is large enough to accept the full scattered beam on out KWS1 and KWS2 SANS defractometer instruments at the 1.8 m detector distance. The larger hole which is < 10 % in area blocked by the mu-metal fingers, will allow wide Q-range measurements up to the maximum allowed by these instruments corresponding to their minimum 1 m detector position. The area shielded by the inner compensation coil in this case will not be important because typical SANS data must use overlapping data sets taken at different detector distances, thus the detector distances can be chosen to cover this gap in Q.

An oven laser and expansion-collimation optics will be prototyped to work with this system. The oven will be heated with either hot flowing air or by electric heaters. The laser will be a very compact 100 W laser diode array bar narrowed with a transmission volume Bragg grating. The collimation optics will include a telescope utilizing a large diameter 45$^\text{o}$ parabolic metal mirror as the last element which will be located inside the magnetic cavity.

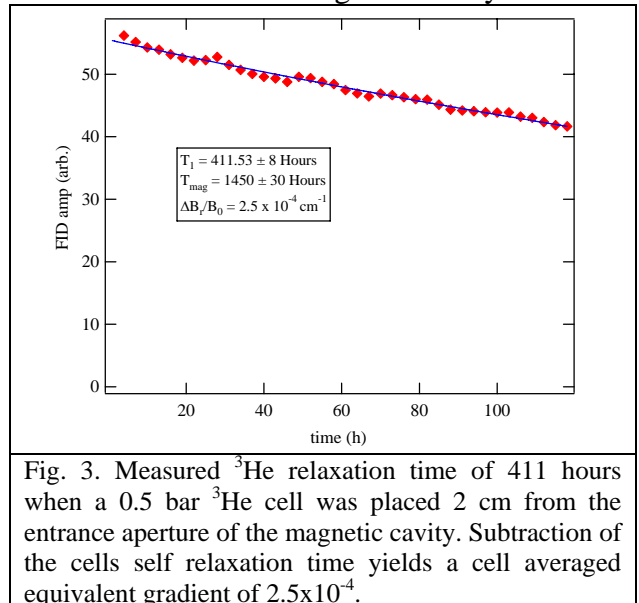

Fig. 3. Measured $^3$He relaxation time of 411 hours when a 0.5 bar $^3$He cell was placed 2 cm from the entrance aperture of the magnetic cavity. Subtraction of the cells self relaxation time yields a cell averaged equivalent gradient of $2.5 \times 10^{-4}$.

The magnetic cavity has been constructed and tested for the polarized $^3$He magnetic lifetime. All of the field coils, main solenoid with 1 turn mm$^{-1}$ and the end-hole compensating coils have been optimized with the number of turns to operate in series from one power supply. Initial experimental optimization was performed with a 3-axis hall probe mounted to a rail to make 1-D longitudinal field maps of the magnetic cavity. The current of the compensating coils was varied to minimize the magnetic field gradients at the desired cell position of 2-8 cm from the inside of the front of the cavity. Then the number of turns in the compensating coils was adjusted for series operation. In the end the final turns did not vary by more than one or two turns from calculations.

After this optimization a 5 cm diameter 5 cm long SEOP cell with 0.5 bar $^3$He pressure and a 930 hour lifetime was polarized externally in our lab and placed in the magnetic cavity to measure its lifetime. It was placed at a distance of 2 cm from the front 1.5x1.5 cm$^2$ hole in the magnetic cavity (right side in Figure 1). Fig. 3 shows an NMR free induction decay measurement of the $^3$He lifetime. After subtracting the contribution of the intrinsic lifetime of the cell which includes $^3$He dipole-dipole relaxation and the wall relaxation of this cell, we obtain a measured $^3$He magnetic lifetime at one bar pressure of 1450 hours or a cell averaged equivalent gradient of $2.5 \times 10^{-4}$ cm$^{-1}$.

This magnetic performance proves that this non-conventional, ultra short solenoid cavity with large openings for scattered beams can provide the required if not excellent performance. Next we will proceed with the prototyping of the volume Bragg grating narrowed lasers, which have already been used successfully in our lab and NIST, and the oven-optics system.

As an example of the expected application, we performed a test experiment on KWS2 using an offline polarized cell. In Fig. 4 we can see the successful subtraction of the incoherent portion of the scattering from the protein sample. This data has been presented and discussed previously in [1,2]

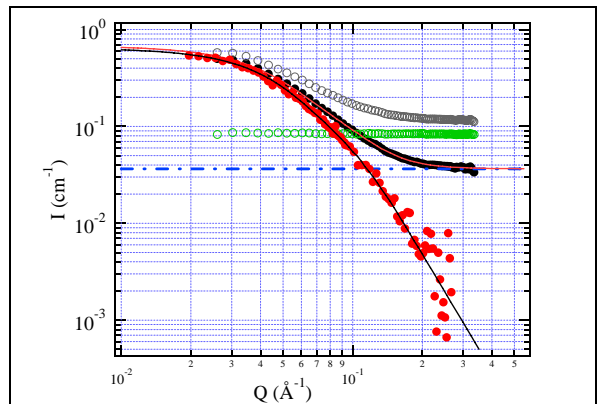

Fig. 4. Sample of the data obtained. The grey/green markers are the data obtained from standard measurements for the sample+solvent and a separate measurement of the solvent respectively. After data treatment one obtains the black/red markers which are the standard SANS signal, and the signal with PA respectively. The red/black lines are fits to the data, and the blue dotted line is the presumed background level from the fit of the standard data. Fitting the PA data to a Beaucage fit with no background, one can obtain the structure dependent fit parameters with no additional assumptions, and the values obtained are consistent with those obtained from standard measurements using several concentrations and no PA. Thus knowledge of the protein is obtained unambiguously with the one measurement employing PA. The data shown is from a sample prepared and by C. Sill who assisted the measurements.